\documentclass[fleqn,twoside]{article}
\usepackage{espcrc2,cite,graphicx,amsmath,amsfonts,amssymb,mathrsfs}
\usepackage[figuresright]{rotating}

\newcommand{\ie}{{\it i.e.}}

\newcommand{\cf}{{\it cf.}}

\newcommand{\eq}{Eq.}

\newcommand{\fig}{Figure}

\newcommand{\Ref}{Ref.}
\newcommand{\Refs}{Refs.}

\newcommand{\stheta}{\sin^22\theta_{13}}
\newcommand{\deltacp}{\delta_{\mathrm{CP}}}

\newcommand{\equ}[1]{\eq~(\ref{equ:#1})}
\newcommand{\figu}[1]{\fig~\ref{fig:#1}}

\newcommand{\bi}{\begin{itemize}}
\newcommand{\ei}{\end{itemize}}

\newcommand{\eet}{\epsilon^m_{e\tau}}
\newcommand{\emt}{\epsilon^m_{\mu\tau}}

\newcommand{\eem}{\epsilon^m_{e\mu}}

\newcommand{\eeta}{|\epsilon^m_{e\tau}|}
\newcommand{\emta}{|\epsilon^m_{\mu\tau}|}

\newcommand{\eema}{|\epsilon^m_{e\mu}|}

\newcommand{\eetp}{\phi^m_{e\tau}}
\newcommand{\emtp}{\phi^m_{\mu\tau}}

\begin{document}

\title{
{\bf Testing non-standard CP violation in neutrino propagation}}

\author{{\large Walter Winter}
\address{{\it Institut f{\"u}r theoretische Physik und Astrophysik, 
Universit{\"a}t W{\"u}rzburg, D-97074 W{\"urzburg}}}\thanks{E-mail: {\tt winter@physik.uni-wuerzburg.de}}}

\begin{abstract}
\noindent {\bf Abstract}
\vspace{2.5mm}

Non-standard physics which can be described by effective four fermion interactions may be an additional source of CP violation in the neutrino propagation. We discuss the detectability of such a CP violation at a neutrino factory. We assume the current baseline setup of the international design study of a neutrino factory (IDS-NF) for the simulation. We find that the CP violation from certain non-standard interactions is, in principle, detectable significantly below their current bounds -- even if there is no CP violation in the standard oscillation framework. Therefore, a new physics effect might be mis-interpreted as the canonical Dirac CP violation, and a possibly even more exciting effect might be missed. 

\vspace*{0.2cm}
\end{abstract}

\maketitle

\section{Introduction}

Physics beyond the Standard Model may introduce non-standard interactions 
(NSI)~\cite{Wolfenstein:1977ue,Valle:1987gv,Guzzo:1991hi,Grossman:1995wx,Roulet:1991sm} 
suppressed by a higher energy scale.  In general, such new physics 
is usually described by effective dimension six~\cite{Buchmuller:1985jz,Bergmann:1998ft,Bergmann:1999pk}
and eight~\cite{Berezhiani:2001rs,Davidson:2003ha} operators. One can describe the
effective dimension $d$ Lagrangian as a function of the non-standard physics scale $\Lambda$ 
as
\begin{equation}
\mathcal{L}^{d} = \lambda \, \frac{\mathcal{O}^d}{\Lambda^{d-4}} \, ,
\label{equ:lageff}
\end{equation}
where $\lambda$ is a dimensionless coupling constant and $\mathcal{O}^d$ is a dimension $d$ 
operator. Thus,
the non-standard physics will be suppressed by $(E_{\mathrm{EWSB}}/\Lambda)^{d-4}$ with
respect to the weak interactions, where $E_{\mathrm{EWSB}}$ is the electroweak symmetry breaking scale.

In this study, we focus on non-standard propagation effects
in standard oscillations (SO). These can be phenomenologically 
described by neutral current-type NSI of the form
\begin{equation}
\mathcal{L}_{\rm NSI}
  =
 \frac{G_{F}}{\sqrt{2}}
 \epsilon^{m}_{\beta \alpha} \, 
 (
 \overline{\nu}_{\beta} \gamma^{\rho} L
 \nu_{\alpha} )
 (
 \bar{f} \gamma_{\rho}  f
 ) + \text{h.c.} \, 
\label{equ:lagrangian}
\end{equation}
with $L=1 - \gamma^5$,
which affect  the neutrino propagation in matter for $f \in \{e, u, d\}$.  Note that, in general, $\epsilon^m_{\alpha \beta}$ are complex numbers for
$\alpha \neq \beta$, and real numbers for $\alpha = \beta$, where we define $\epsilon^m_{\alpha \beta} \equiv|\epsilon^m_{\alpha \beta}| \exp( i \phi^m_{\alpha \beta})$. Thus, $\eem$, $\emt$, $\eet$ are possible sources of non-standard CP violation (NSI-CPV).\footnote{In \Refs~\cite{Gonzalez-Garcia:2001mp,FernandezMartinez:2007ms} such NSI-CPV was discussed in the context of source and detection NSI, whereas we focus on the propagation effects. Note that, depending on the model, source and propagation NSI could be related.
However, the simplest allowed models to induce $\emt$ or $\eet$ involve two mediator fields (and some cancellation conditions), and propagation NSI are not related to source and detection NSI~\cite{Gavela:2008ra}. } 

They enter the propagation Hamiltonian in flavor base proportional to the matter potential $a_{\mathrm{CC}}=2 \sqrt{2} E G_F N_e$ (with $N_e$ the electron density) in the off-diagonal elements.
Since $\eema$ is very well constrained, we focus on $\emta$ and $\eeta$, for which the current bounds are $\mathcal{O}(0.1)$ and $\mathcal{O}(1)$, respectively~\cite{Davidson:2003ha,Barranco:2007ej,GonzalezGarcia:2007ib}. Therefore, the phases $\emtp$ and $\eetp$ might be accessible by future experiments for large enough $\emta$ and $\eeta$. 
The necessary conditions for an underlying model producing such large NSI are discussed elsewhere~\cite{Gavela:2008ra,Antusch:2008tz}.
Since these interactions will be suppressed by at least a factor of $\Lambda^2$ (\cf, \equ{lageff}), it might be plausible to look for NSI-CPV in the best discussed neutrino oscillation experiments which are sensitive to the highest $\Lambda$-scales, such as neutrino factories~\cite{Geer:1998iz,Barger:1999fs}.

The measurement of NSIs in neutrino factories has been discussed in \Refs~\cite{Huber:2001zw,Gonzalez-Garcia:2001mp,Gago:2001xg,Huber:2002bi,Blennow:2005qj,Kopp:2007mi,Ribeiro:2007ud,Kopp:2008ds}. As illustrated in \Refs~\cite{Ribeiro:2007ud,Kopp:2008ds}, the disappearance channels and second ``magic'' baseline~\cite{Huber:2003ak} are mandatory for excellent NSI sensitivities. 
In particular, $\emt$ is best measured with the disappearance channel, whereas $\eet$ is best measured with the appearance channel. Therefore, we expect that the measurement of  $\eetp$ will be qualitatively similar to that of $\deltacp$, whereas that of $\emtp$ will have completely new characteristics. We use in this study the baseline setup of the international design study for a neutrino factory~\cite{ids}, which includes two baselines, as well as the disappearance channels by the measurement of the wrong-sign muons. 

A focus of this letter is to demonstrate that the discovery of NSI-CPV
should be quantified with performance indicators similar to SO-CPV. 
In addition, the full (relevant) parameter space using a full simulation is discussed. 
Since there is no model-independent connection between source or detection and propagation NSI~\cite{Gavela:2008ra}
and there is not yet any near detector specification for the neutrino factory within the
international design study, we do not discuss source and propagation NSI.

\section{Method and performance indicator}

Our simulations use the GLoBES software~\cite{Huber:2004ka,Huber:2007ji} with the current best-fit values and solar oscillation parameter uncertainties from \Ref~\cite{GonzalezGarcia:2007ib}, as well as a 2\% error on the (constant) matter density profile, \ie, we expect the matter density profile to be known with that precision.\footnote{In fact, we have checked that the impact of a larger matter density uncertainty on the $\eet$ measurement is very small, at the level of a few percent correction.}
 For the sake of simplicity, we use a normal simulated mass hierarchy.  The experimental scenario we consider is the {\sf IDS-NF}~1.0 setup from \Ref~\cite{ids}, which is the current standard setup for the ``International design study of the neutrino factory'' ({\sf IDS-NF}). This setup has been optimized within \Refs~\cite{Huber:2006wb,Bandyopadhyay:2007kx} for the measurement of $\stheta$, the neutrino mass hierarchy, and leptonic CP violation in the case of standard oscillations. 
In short, it uses two baselines at about $4 \, 000 \, \mathrm{km}$ and $7 \, 500 \, \mathrm{km}$ with two (identical) magnetized iron neutrino detectors (MIND) with a fiducial mass of $50 \, \mathrm{kt}$ each.  For each baseline, a total of $2.5 \cdot 10^{21}$ useful muon decays plus $2.5 \cdot 10^{21}$ useful anti-muon decays in the straight of the corresponding storage ring is used, which could be achieved by ten years of operation with $2.5 \cdot 10^{20}$ useful muon decays per baseline, year, and polarity. The muon energy $E_\mu$ is assumed to be $25 \, \mathrm{GeV}$, which is sufficient for a detector with a low enough detection threshold~\cite{Huber:2006wb}. The detector and systematics specifications can be found in \Refs~\cite{ids}. Note that there is not yet any near detector specification. We do not simulate the near detector explicitely, because we do not discuss non-standard production or detection effects such as in \Ref~\cite{Kopp:2007ne}.  As a small modification of the {\sf IDS-NF} baseline setup, we do not include the emulsion cloud chamber for $\nu_\tau$ detection at the short baseline, since it has been demonstrated in \Ref~\cite{Kopp:2008ds} that it hardly contributes to the SO and NSI sensitivities if two baselines are used. We have checked that this also applies for a (hypothetical) $\nu_\mu \rightarrow \nu_\tau$ oscillation channel for the effects discussed in this study (which might be different for NSIs in the production process, see \Ref~\cite{FernandezMartinez:2007ms}).

\begin{figure}[t!]
\begin{center}
\includegraphics[width=0.85\columnwidth]{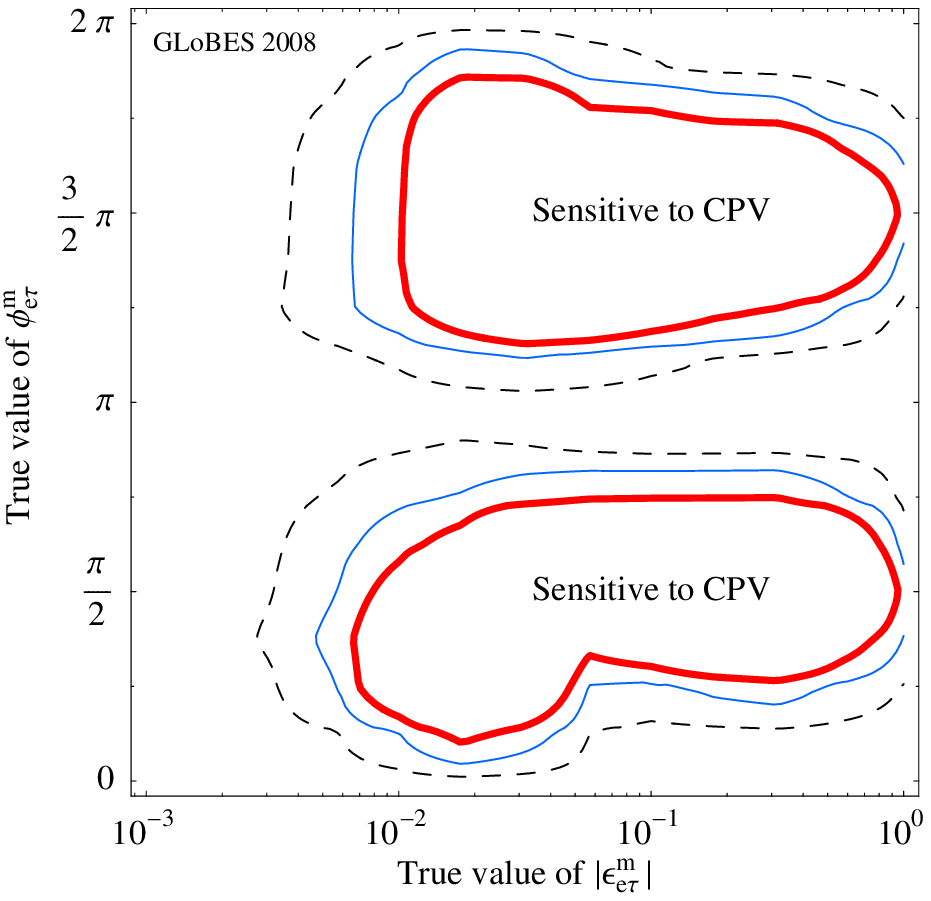}
\vspace*{-1cm}
\end{center}
\caption{\label{fig:did} Sensitivity to NSI-CPV in $\eet$ as a function of the true values of $\eeta$ and $\eetp$ (for the true $\stheta=0$). The different contours correspond to $\Delta \chi^2=1$ (dashed), $4$ (thin solid), and $9$ (thick).
}
\end{figure}

We define the sensitivity to NSI-CPV, in the same way as the sensitivity to SO-CPV, as the $\Delta \chi^2$ with which any CP conserving solution can be excluded. 
That is, we simulate a true $\epsilon^m_{\alpha \beta}=|\epsilon^m_{\alpha \beta}| \, \exp (i \phi^m_{\alpha \beta})$,
where $\phi^m_{\alpha \beta}$ is the CP violating phase $\notin \{0 , \pi\}$. In addition, we have a set of simulated values for the SO parameters. Then we compute the $\Delta \chi^2$ for $\phi^m_{\alpha \beta}$ (fit) fixed to $0$ and $\pi$ (CP conservation) and choose the minimum between these two values. All the fit SO parameters and $| \epsilon^m_{\alpha \beta}|$ (fit) are marginalized over. For the sake of simplicity, we do not take into account the mass hierarchy degeneracy.

\section{Discovery of non-standard CP violation}

\begin{figure*}[t!]
\begin{center}
\includegraphics[width=0.9\textwidth]{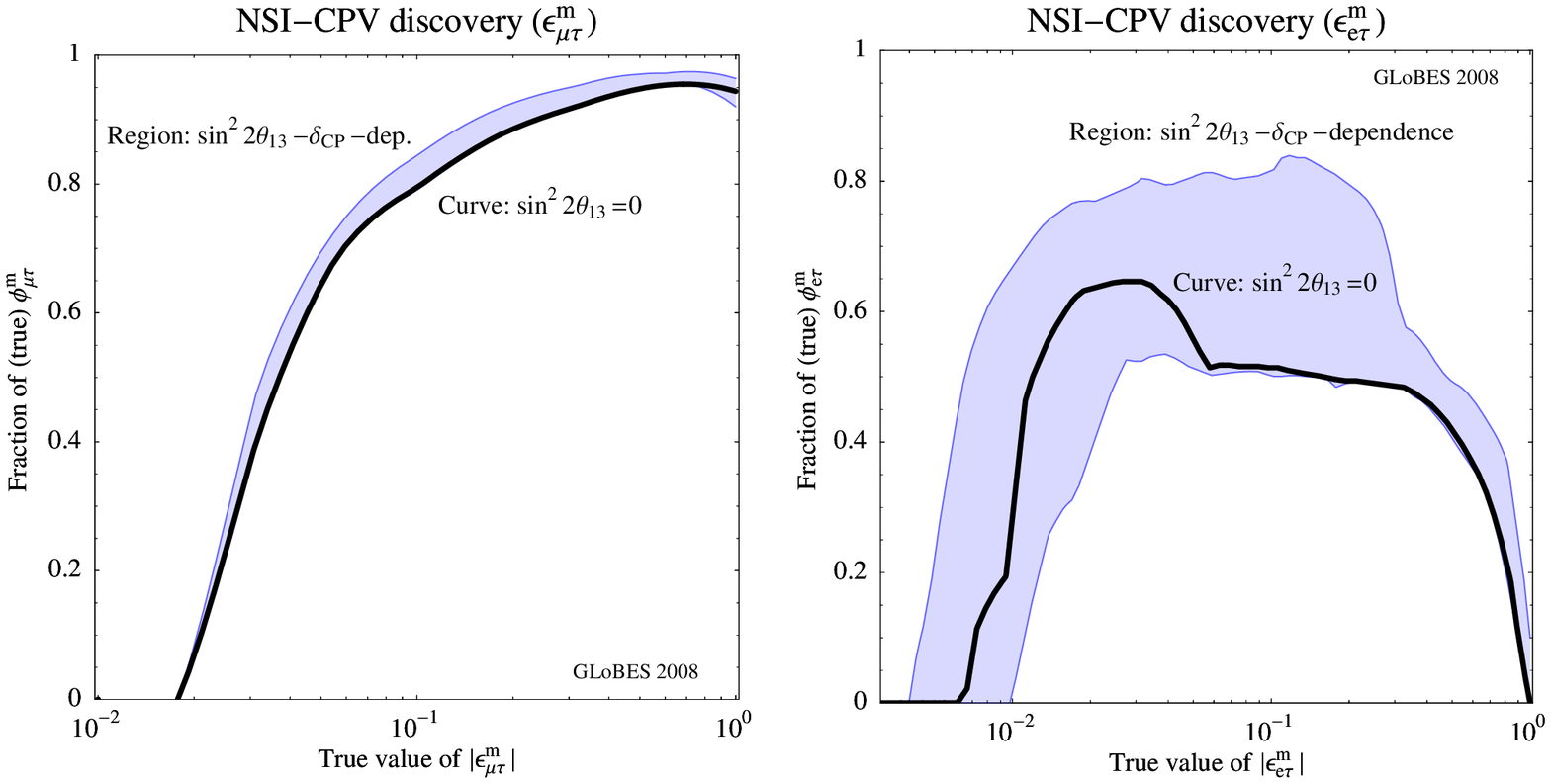}
\vspace*{-1cm}
\end{center}
\caption{\label{fig:nscpv}Fraction of $\emtp$ (left) and $\eetp$ (right) for which non-standard CP violation
will be discovered (at $\Delta \chi^2 = 9$) as a function of $\emta$ (left) and $\eeta$ (right). The thick curves are computed for $\stheta=0$. The shaded regions illustrate the dependence on the (true) $\stheta$ and $\deltacp$.
}
\end{figure*}
In the context of NSI-CPV, the most important question might be for which region of the parameter space
NSI-CPV will be discovered at a neutrino factory. Since the solar and atmospheric oscillation parameters are very well known, the performance will mainly depend on the true values of $|\epsilon^m_{\alpha \beta}|$, $\phi^m_{\alpha \beta}$, $\stheta$, and $\deltacp$, where $\phi^m_{\alpha \beta}$ describes the CP violation of interest. 
Because the absolute value of $\epsilon^m_{\alpha \beta}$ suppresses the phase measurement, the simulated $| \epsilon^m_{\alpha \beta} |$ and $\phi^m_{\alpha \beta}$ will be the most important parameters for the parameter space test, similar to $\stheta$ and $\deltacp$ for the SO-CPV. We illustrate this dependence in \figu{did} for $\eet$ as a function of the true values of $\eeta$ and $\eetp$, as well as the true $\stheta=0$. This figure looks very similar to the corresponding SO-CPV figure as a function of $\stheta$ and $\deltacp$: There is a cutoff at small $\eeta$, below which the phase effects are suppressed, and there is no sensitivity close to the CP-conserving solutions $\eetp=0$ and $\pi$. Therefore, we adopt an approach similar to that of SO-CPV. We show in \figu{nscpv} the fraction of $\emtp$ (left) and $\eetp$ (right) for which NSI-CPV will be discovered  as a function of the $\emta$ (left) and $\eeta$ (right). In this case, the fraction of $\epsilon_{\alpha \beta}^m$ represents the stacking of all sensitive regions in \figu{did} along any vertical line corresponding to any fixed $\eeta$. In \figu{nscpv}, the dependence on the true $\stheta$ and $\deltacp$ is indicated by the shaded regions, whereas the curves correspond to the true $\stheta=0$ (\ie, the thick curve in the right panel corresponds to the thick curve in \figu{did}). 

As it is obvious from the analytical and quantitative discussion in \Ref~\cite{Kopp:2008ds}, the $\nu_e \rightarrow \nu_\mu$ (and $\bar\nu_e \rightarrow \bar\nu_\mu$) appearance channels will dominate the determination of $\eetp$, whereas the $\nu_\mu \rightarrow \nu_\mu$ (and $\bar\nu_\mu \rightarrow \bar\nu_\mu$) disappearance channels will dominate the the determination of $\emtp$ (see also \Ref~\cite{Campanelli:2002cc} for more analytical discussions). For that reason, we obtain a strong dependence on the simulated $\stheta$ and $\deltacp$ for $\eetp$ (right panel of \figu{nscpv}), because the appearance channels are most sensitive to these SO parameters, whereas the CP violation in $\emtp$ is hardly affected by these parameters. 
From \figu{nscpv} (left panel), we can read off that NSI-CPV will be discovered for about 80\% of all possible $\emtp$ for $\emta \sim 0.1$ close to the current bound. The $\emta$ reach is, however, limited to $\emta \gtrsim 0.02$, which means that any significant improvement of the bound will exclude this possibility. 
For $\eetp$ in the right panel, we obtain a picture qualitatively similar to the SO-CPV measurement because of the dominance of the appearance channels.
We obtain a large fraction of $\eetp$ of up to 80\% in an intermediate range $0.01 \lesssim \eeta \lesssim 0.3$.
From the analytical discussion in \Ref~\cite{Kopp:2008ds} (Eqs. (10) and (11)),  we cannot observe the NSI-CPV for too large $\eeta$, because the terms quadratic in $\eeta$ will then be too large a background for $\eeta \gtrsim \mathcal{O}(1)$. This means that although the NSI can be easily established, the NSI-CPV cannot be established against the phase-independent terms.

\section{Mis-interpretation of NSI-CPV}

\begin{figure*}[t!]
\begin{center}
\includegraphics[height=6cm]{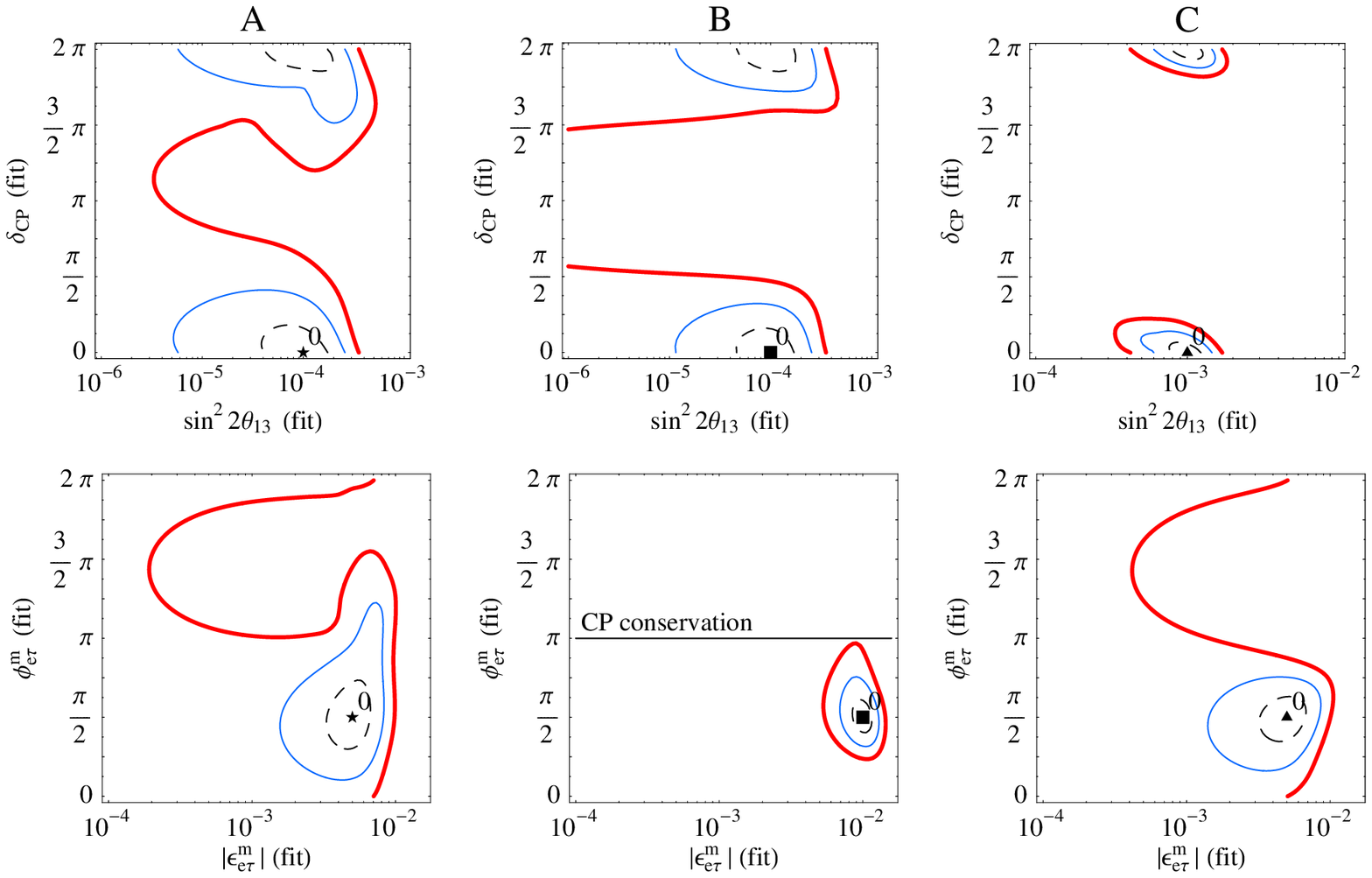}
\includegraphics[height=6cm]{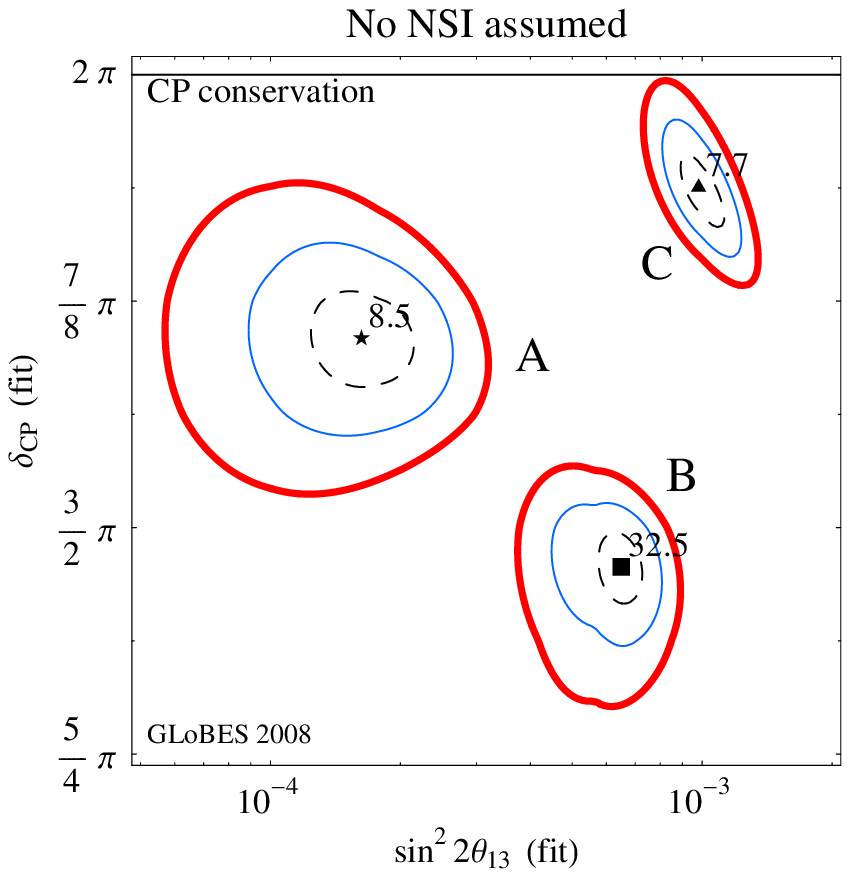}
\vspace*{-1cm}
\end{center}
\caption{\label{fig:wrongfit} NSI-CPV misinterpreted as SO-CPV.
The first three columns represent the fits in the $\stheta$-$\deltacp$ (upper row) and $\eeta$-$\eetp$ (lower row) planes including NSI (from $\eet$) for three different sets of true values: case~A with $\stheta=10^{-4}$ and $\eeta=0.005$, case~B with $\stheta=10^{-4}$ and $\eeta=0.01$, and case~C with $\stheta=10^{-3}$ and $\eeta=0.005$.  In all panels, we choose the true $\deltacp=0$ (no SO-CPV) and the true $\eetp=\pi/4$ (maximal NSI-CPV). In the right panel, we show the same fits in the $\stheta$-$\deltacp$ plane assuming standard oscillations only (no NSI).  In all panels, the contours represent $\Delta \chi^2=1$ (dashed), $4$ (solid thin), and $9$ (solid thick), and the best-fit values are marked by symbols (including the minimum $\chi^2$).
}
\end{figure*}
Let us assume that there is no SO-CPV in nature, but there is large NSI-CPV.
What would such a fit look like, and when would one confuse the
NSI-CPV with the SO-CPV? Since $\emtp$ is measured in the disappearance channel, which is hardly affected by $\deltacp$, we focus on $\eetp$ in this section.

As an example, we choose a true $\deltacp=0$ (CP conservation) and the true $\eetp=\pi/2$ (maximal CP violation), and we simulate three specific sets of true values:
\begin{description}
\item[A] $\stheta=10^{-4}$, $\eeta=0.005$
\item[B] $\stheta=10^{-4}$, $\eeta=0.01$
\item[C] $\stheta=10^{-3}$, $\eeta=0.005$
\end{description}
There is some parameter dependence in these choices, but the examples are good enough to illustrate the qualitative main points. 
We show in \figu{wrongfit}, left three columns, the fits in the $\stheta$-$\deltacp$ (upper row) and $\eeta$-$\eetp$ (lower row) planes assuming the NSI scenario including $\eet$. Obviously, in all cases, the minimal $\chi^2=0$ because we simulate the average experiment performance without statistical fluctuations. At $\Delta \chi^2=9$, NSI-CPV can neither be established in case~A nor in case~C, whereas it can be measured in case~B (lower row). On the other hand, $\deltacp$ can be measured in cases~B, and~C, whereas no information can be obtained in case~A (upper row). Generally speaking, the true $\stheta$ or true $\eeta$ have to be large enough to observe the corresponding phase.
In the right panel of \figu{wrongfit}, we illustrate the effect of the wrong hypothesis: If only standard oscillations are assumed (and $\eeta=0$ in the fit), the minimal $\chi^2$ in all three cases will be non-vanishing, and might be confused with statistical fluctuations. In all three cases, the simulated maximal NSI-CPV will be mis-interpreted as SO-CPV, because the CP-conserving values $\deltacp=0$ and $\pi$ can be excluded. Note that especially in cases~A and~C the minimal $\chi^2$ is relatively small, and the confusion might not be obvious. Even worse, in these cases neither SO-CPV nor NSI-CPV would be established in the correct fit.

\section{Sensitivity to {\em any} CP violation}

\begin{figure}[t!]
\begin{center}
\includegraphics[width=0.85\columnwidth]{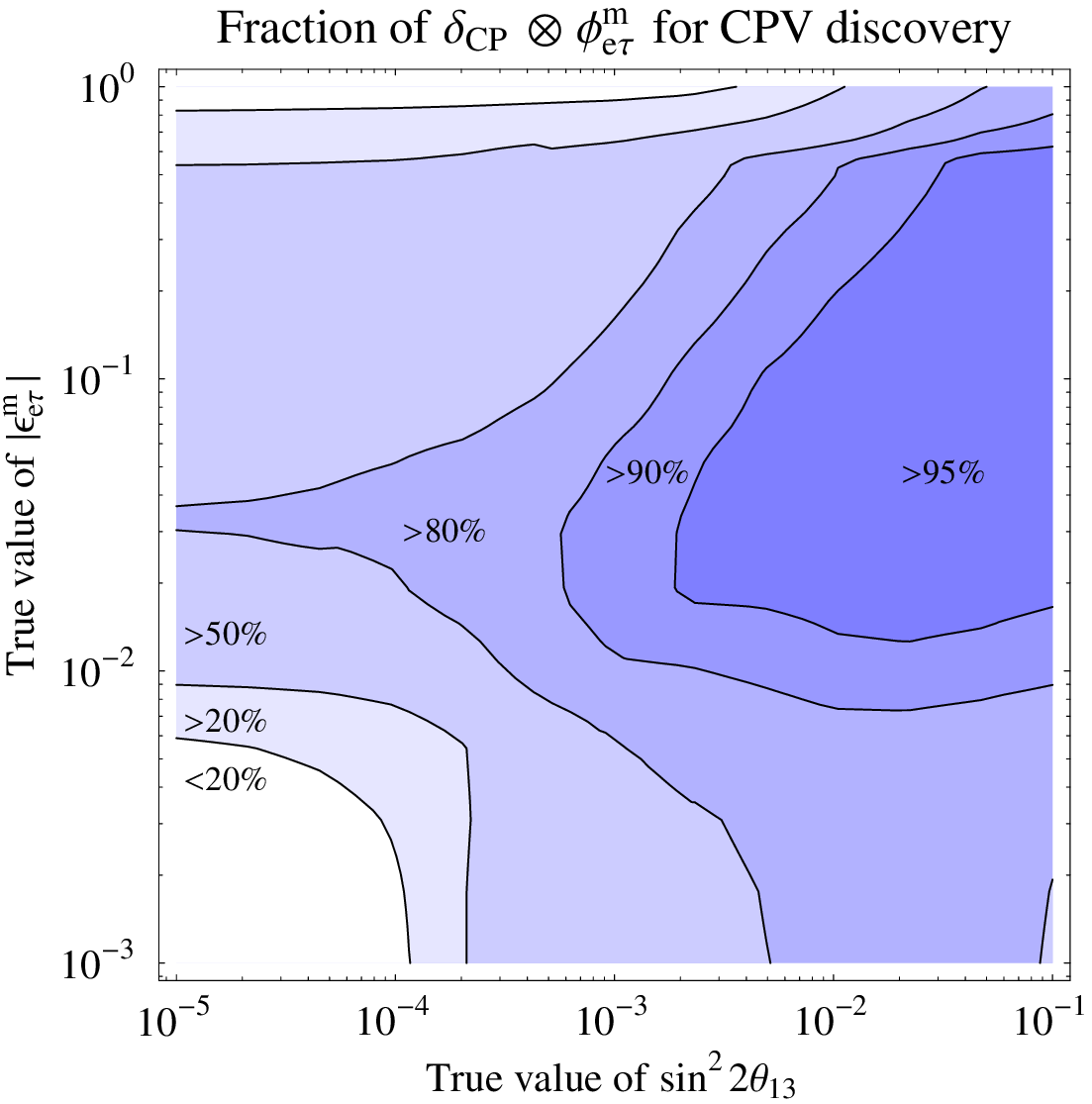}
\vspace*{-1cm}
\end{center}
\caption{\label{fig:anycpv} Sensitivity to any (SO or NSI) CPV as a function of the true $\stheta$ and true $\eeta$. The sensitivity is given as a fraction of all (true) $\deltacp \otimes \eetp$ for which CPV can be established ($\Delta \chi^2=9$).
}
\end{figure}
Let us suppose that there are NSI, and there might be CP violation from either $\deltacp$ or $\eetp$.
In this case, it may be an interesting question if any CP violation can be established, no matter of the origin. This question is equivalent to excluding the CP conserving solutions $(\deltacp,\eetp) \in \{(0, 0), (0, \pi), (\pi, 0) , (\pi,\pi) \}$. 
Similar to the ``fraction of $\deltacp$'' or ``fraction of $\eetp$'' for the single parameter measurement, we define, for a given set of true $\stheta$ and $\eeta$, the fraction of (true) $\deltacp \otimes \eetp$ as the fraction of the $\deltacp \otimes \eetp$ plane for which (any) CP violation can be established. This means that for a random pick of $\deltacp$ and $\eetp$ (uniformly distributed in the phases), this parameter will tell the probability that CP violation can be established. We show this sensitivity to any (SO or NSI) CPV as a function of the true $\stheta$ and true $\eeta$
in \figu{anycpv}. 
Obviously, if both $\stheta$ and $\eeta$ are small, no CPV can be established. If only one of the parameters is small, the sensitivity corresponds to the single parameter measurement (for the other parameter and its phase being marginalized over). If both $\stheta$ and $\eeta$ are large and in their optimal ranges, the fraction of $\deltacp \otimes \eetp$ can reach more than 95\%. In this case, CPV cannot be established in only very small regions around the four CP-conserving solutions given above. In the probability interpretation, the chance to find CPV is therefore, in fact, larger if both large $\stheta$ and $\eeta$ are present, compared to the case of no NSI.

\section{Summary and conclusions}

We have demonstrated that, in a neutrino factory corresponding to the current IDS-NF baseline setup, CP violation from non-standard neutrino interactions in matter could be measured if $\emta \gtrsim 0.02$ or $\eeta \gtrsim 0.005$.
This observation is almost independent of the true values of $\stheta$ and $\deltacp$. We have also shown that there are regions in parameter space where an erroneous assumption of standard oscillations only will lead to a fake CP violation signal in the presence of NSI, even if CP is conserved in $\deltacp$. Finally, we have discussed the chances to measure {\em any} CP violation in the presence of both large enough $\stheta$ and $\eeta$. We have found that CP violation will be found in up to 95\% of all possible phase combinations for $\deltacp$ and $\eetp$. For the quantification, we have used performance indicators similar to the standard CP violation measurement.

We conclude that even if there is no CP violation in standard oscillations or $\stheta$ is too small to detect it, a neutrino factory has the chance to find CP violation from new physics effects. In any future analysis, it is therefore important to carefully consider the possibility of non-standard effects, in order not to overlook an even more   interesting hint for new physics.

{\bf Acknowledgments:}
I would like to thank Joachim Kopp and Toshihiko Ota for providing 
the NSI implementation into GLoBES, and for useful discussions and comments.
In addition, I would like to thank Michele Maltoni for some statistics
advice.
This work has been supported by the Emmy Noether program of
DFG.

\vspace*{-0.3cm}

{\small

}

\end{document}